\documentclass[a4paper,12pt]{article}
\pdfoutput=1
\usepackage{jheppub}
\usepackage{hyperref}
\hypersetup{colorlinks=true}
\usepackage[T1]{fontenc} 
\title{Neumann-Rosochatius system for (m,n) string in $AdS_3 \times S^3$ with mixed flux}
\author[a]{Adrita Chakraborty}
\author[b]{~{\rm and} Kamal L. Panigrahi}
 \affiliation[a]{Centre For Theoretical Studies, Indian Institute of Technology Kharagpur-721302, India}
\affiliation[b]{Department of Physics, Indian Institute of Technology Kharagpur-721302, India}
\emailAdd{adimanta09@iitkgp.ac.in}
\emailAdd{panigrahi@phy.iitkgp.ac.in}
\abstract{$SL(2,\mathbb{Z})$ invariant action for probe  $(m,n)$ string in $AdS_3\times S^3\times T^4$ with mixed three-form fluxes can be described by an integrable deformation of an one-dimensional Neumann-Rosochatius (NR) system. We present the deformed features of the integrable model and study general class of rotating and pulsating solutions by solving the integrable equations of motion.  For the rotating string, the explicit solutions can be expressed in terms of elliptic functions. We make use of the integrals of motion to find out the scaling relation among conserved charges for the particular case of constant radii solutions. Then we study the closed $(m,n)$ string pulsating in $R_t\times S^3$. We find the string profile and calculate the total energy of such pulsating string in terms of oscillation number $(\cal{N})$ and angular momentum $(\cal{J})$.}

\begin{document}
\maketitle
	\flushbottom
	\section{Introduction}	
Planar integrability on both sides of the famous AdS/CFT correspondence \cite{Maldacena:1997re,Gubser:1998bc,Witten:1998qj} has been proved to be very promising technique to unravel a deeper understanding in the study of string spectrum in different semisymmetric superspaces \cite{Zarembo:2010sg}. Minahan and Zarembo in \cite{Minahan:2002ve} first established the matching between the one-loop dilatation operators of the SU(2) sector of $\mathcal{N}=4$ SYM theory with the Hamiltonian of integrable SO(6) spin chain model \cite{Beisert:2003jj,Beisert:2003yb}. In proving the AdS/CFT duality in the large charge limit, varieties of rigidly rotating strings have been studied in different backgrounds to enhance our understanding of the dual states in the gravity side. Among them special cases on sphere consist of giant magnon \cite{Hofman:2006xt} and spiky string \cite{Kruczenski:2004wg} solutions which are dual to elementary excitations with large momentum $p$ and higher twist operators respectively in the field theory side of the duality. Also in \cite{Minahan:2002rc} an exact correspondence was proposed between string states and some dual field theory operators by considering spinning as well as pulsating string solutions, the later  first developed in \cite{Gubser:2002tv}. Despite being more stable solutions \cite{Khan:2005fc} these types of string state have not been explored enough as compared to the rigidly rotating strings. Pulsating string solutions were explored earlier in  $AdS_5\times S^5$ background  \cite{Khan:2003sm,Smedback:1998yn} and then in  $AdS_4\times \mathbb{CP}^3$  \cite{Chen:2008qq,Dimov:2009rd}. In some prior works \cite{Engquist:2003rn,Beisert:2003xu,Beisert:2003ea} the YM duals for semiclassical spinning and pulsating string states in $AdS_5\times S^5$ were analysed in terms of integrable spin chain description of the field theory. Over the last few years, $AdS_{3}\times S^{3}\times M^{4}$ geometry supported by mixed three form fluxes (both NS-NS and R-R) has been studied as a nice tool in the context of classical integrability in $AdS_{3}/CFT_{2}$ correspondence \cite{Cagnazzo:2012se,Hoare:2013lja,Hoare:2013ida,Hoare:2013pma,Pittelli:2014ria,Pittelli:2017spf,Baggio:2018gct}which is a much explored holographic extension of the $AdS_5/CFT_4$ duality proposed earlier. The appearance of integrability as a symmetry in Green-Schwarz action of type IIB string in compactified $AdS_{3}\times S^{3}\times M^{4}$  background with mixed R-R and NS-NS three form flux has put forth a renewed interest to analyse the $AdS_{3}/CFT_{2}$ duality in the presence of mixed fluxes by means of well known classical integrable models. Though the explicit form of the dual theory is yet to explore, such background seems to be a promising laboratory to perform various general string solutions, consequently giving some ideas of the respective dual operators.\\

	A general description of the finite gap solutions of classically integrable string sigma model has been demonstrated in \cite{Kazakov:2004qf,Beisert:2004ag} in terms of solutions of certain integrable models. In this regard, it is worth emphasizing that a generalised ansatz for string coordinates \cite{Kruczenski:2006pk,Arutyunov:2003uj,Arutyunov:2003za} provides a method to reformulate large class of string configurations in terms of solutions of very well-known one dimensional integrable Neumann or Neumann-Rosochatius (NR) system. The Neumann integrable model describes harmonic oscillator restricted to move on a sphere whereas the NR system is an integrable extension of the former with an additional centrifugal potential barrier of the form $\frac{1}{r^{2}}$. These mechanical models have been proved to be very effective to deal with the problem of geodesics on ellipsoid or equivariant harmonic maps into sphere \cite{Uhlenbeck,Babelon:2003qtg,Avan:1991ib,Avan:1989dn}. Earlier, in the context of classical integrability, classical giant magnon \cite{Kruczenski:2006pk,Arutyunov:2003uj,Arutyunov:2003za} and spiky string \cite{Ishizeki:2007we,Mosaffa:2007ty,Hayashi:2007bq} solutions on $AdS_{5}\times S^{5}$ were obtained by using NR model approach. Moreover, there are several recent works \cite{Ahn:2008hj,Hernandez:2014eta,Hernandez:2015nba,Arutyunov:2016ysi,Hernandez:2017raj,Hernandez:2018gcd,Chakraborty:2019gmt} which reveal that general solutions corresponding to closed rotating and pulsating string in various 10D AdS/S compact spaces with pure or mixed flux may be constructed from systematic analysis of the one dimensional Lagrangian of NR system with integrable deformation. 

In a seminal paper, J. H. Schwarz  \cite{Schwarz:1995dk} had constructed an $SL(2, {\mathbb Z})$ multiplet of string-like solutions in type IIB supergravity starting from the fundamental string solution. It is also known that the equations of motion of type IIB theory are invariant under an $SL(2, {\mathbb R})$ rotation group. This enables one to generate new supergravity solutions in type IIB theory by applying this rotation to known solutions in supergravity. A discrete subgroup 
$SL(2, {\mathbb Z})$ of this $SL(2, {\mathbb R})$ group has been conjectured later to be the exact symmetry group of the type IIB string theory. Till date, a lot of work has been devoted for constructing various string-like as well as five-brane solutions of type IIB supergravity equations using $SL(2,\mathbb{Z})$ invariance of low energy effective action of this theory \cite{Schwarz:1995dk,Tseytlin:1996it,Lu:1998vh,Lozano:1997cy,Cederwall:1997ts,Kluson:2016dca,Kluson:2016pxg}. 
In the near horizon limit, an $SL(2, {\mathbb Z})$ transformed bound state solution of $Q_5$ NS5-branes and $Q_1$ F-strings, gives rise to $AdS_3\times S^3$ background with mixed three form fluxes having integer charges. It has also been recently shown that the $SL(2,{\mathbb{Z}})$-transformation and the near horizon limit commute. This allows us to map the $(m,n)$ string action in $AdS_{3} \times S^{3}$ background with mixed three form fluxes to $(m',n')$ string action in $AdS_{3}\times S^{3}$ background with NS-NS two form flux using the relation
\begin{equation}
    \textbf{m}'=\left(\begin{array}{c} m'\\n'\end{array}\right)=\left(\begin{array}{cc}d&-b\\-c&a\end{array}\right)\left(\begin{array}{c} m\\n\end{array}\right),
\end{equation} where $a,b,c$ and $d$ are the integer entries of $SL(2,\mathbb{Z})$ matrix with $ad-bc=1$. Interesting analysis of this fact has so far been put forth with $(m,n)$ string action as natural probe in various backgrounds \cite{Kluson:2016dca,Kluson:2016pxg,Banerjee:2016avv,Barik:2016yav,Kluson:2019ifd}. In \cite{Barik:2016yav}, some fascinating observations on giant magnon and single spike solutions for rigidly rotating string along with the analysis of pulsating string as  probe $(m,n)$ string in mixed flux background have been performed by mapping it into $(m',n')$ string in $AdS_{3}\times S^{3}$ background with NSNS two form flux.

Motivated by such analysis, we cast our focus on the manifest $SL(2,\mathbb{Z})$ covariant action of probe $(m,n)$ string in $AdS_{3}\times S^{3}\times T^{4}$ background considering the presence of both the three form NSNS and RR fluxes using general rotating and pulsating string ansatz in the light of classical integrability of both sides of the correspondence. In this article we wish to construct an integrable deformed NR system and utilize the solutions of integrable equations of motion of that system to obtain the relations between conserved charges for both rotating and pulsating  $(m,n)$ string. The rest of the paper is organized as follows. 
In section 2,  we present the construction of deformed Lagrangian and Hamiltonian of integrable NR model for closed  $(m,n)$-string rotating in $R_{t}\times S^{3}$ background in the presence of mixed three form fluxes. Also we discuss for corresponding Uhlenbeck integrals of motion for such system. We derive elliptic solutions for rotating $(m,n)$ string which resemble the circular-type string solutions. We further study of conserved charges and energy-angular momenta dispersion relation for constant radii case for the closed rotating string. Section 4 is devoted to the study of $(m,n)$ type pulsating string in mixed flux background from the solutions of the modified NR model. We study the string profile and compute small energy correction to the scaling relation in terms of oscillation number in short string limit. Finally, we conclude our results in section 4. 
\section{Modified NR model for rotating $(m,n)$ string in $AdS_{3}\times S^{3}$ with mixed flux}
We start by writing down the background metric and associated NS-NS and R-R fluxes for $AdS_{3}\times S^{3}$ background 
\begin{subequations}
   \begin{align}
&ds^{2}=L^{2}\left[ds^{2}_{\tilde{AdS}_{3}}+ds^{2}_{S^{3}}\right]\label{line element1} \\&
H=2L^{2}\left(\tilde{\epsilon}_{AdS_{3}}+\epsilon_{S^{3} }\right), ~~L^{2}=r_{5}^{2}\\&
e^{-2\phi}=\frac{r_{1}^{2}}{g_{s}^{2}r_{5}^{2}},~~r_{1}^{2}=\frac{16\pi^{4}g_{s}^{2}\alpha^{'3}Q_{1}}{V_{4}},~~r_{5}^{2}=Q_{5}\alpha^{'},~~V_{4}=(2\pi)^{4}\alpha^{'2}v
 \end{align}
\end{subequations}Here $ds^{2}_{\tilde{AdS_{3}}}$ is the line element of the corresponding background in terms of dimensionless variables. 
In this section, we wish to study the dynamics of $(m,n)$ string in general background with mixed three form flux . 
On the other hand, the action of $(m,n)$ string in general background with metric $g_{MN}$ and flux ${\bf B}_{MN}$ is given by
\begin{eqnarray}
S_{m,n} = -T_{D1}\int d\tau d\sigma\left[\sqrt{\textbf{m}^{T}\mathcal{M}^{-1}\textbf{m}}\sqrt{-\det g_{MN}\partial_{\alpha}x^{M}\partial_{\beta}x^{N}} -  \epsilon^{\alpha\beta}\textbf{m}^{T}\textbf{B}_{MN}\partial_{\alpha}x^{M}\partial_{\beta}x^{N}\right],
 \nonumber \\
        \label{action1}
    \end{eqnarray}with $g_{MN}$ to be the Einstein frame metric which is invariant under $SL(2,\mathbb{Z})$ transformations. This action was derived in \cite{Kluson:2016pxg} by applying $SL(2,\mathbb{Z})$ invariance of the action of $n$ coincident D1-branes in general background. The vector $\textbf{B}$ is defined as 
    \begin{equation}
        \textbf{B}=\left(\begin{array}{c} B^{(2)}\\C^{(2)}\end{array}\right)
        \label{equationB}
    \end{equation}where $B^{(2)}$ and $C^{(2)}$ are NS-NS and R-R two form fluxes respectively. It transforms under $SL(2,\mathbb{Z})$ transformation as
    \begin{equation}
        \hat{\textbf{B}}=(\Lambda^{T})^{-1}\textbf{B}
        \label{SL2ZB}
    \end{equation}where $\Lambda$ is given by
       \begin{equation}
     \Lambda=\left(\begin{array}{cc}a&b\\c&d\end{array}\right),~~~{\rm with}~~~~ad-bc=1
     \label{equationlambda}
\end{equation} It is convenient to introduce a complex field $\tau=\chi +ie^{-\phi}$ containing both the axion scalar $\chi$ and dilaton $\phi$ corresponding to NS-NS and R-R sectors respectively. With this complex field, we introduce a matrix in the following form:
    \begin{equation}
        \mathcal{M}=e^{\phi}\left(\begin{array}{cc}\tau\tau^{*}&\chi\\\chi&1\end{array}\right)=e^{\phi}\left(\begin{array}{cc}\chi^{2}+e^{-2\phi}&\chi\\\chi&1\end{array}\right),~~\det\mathcal{M}=1
        \label{equationM}
    \end{equation}which transforms under $SL(2,\mathbb{Z})$ transformation as,
    \begin{equation}
    \hat{\mathcal{M}}=\Lambda\mathcal{M}\Lambda^{T} \ .
    \label{SL2ZM}
        \end{equation} 
On the other hand, the matrix $\textbf{m}$ in the action (\ref{action1}) is expressed as
\begin{equation}
 \textbf{m}=\left(\begin{array}{c} m\\n\end{array}\right)
 \label{equationform}
\end{equation}
where $m$ and $n$ are the integers counting for the number of fundamental strings coupled to the NS-NS flux and D1-branes coupled to the R-R flux respectively. It transforms under $SL(2,\mathbb{Z})$ as $\hat{\textbf{m}}=\Lambda\textbf{m}$ whereas the $SL(2,\mathbb{Z})$ transformation for $B$ and $\mathcal{M}$ reads as equations (\ref{SL2ZB}) and (\ref{SL2ZM}) respectively. Hence this action becomes manifestly $SL(2,\mathbb{Z})$ invariant. Using (\ref{equationB}) and (\ref{equationM}) we have
        \begin{equation*}
              \textbf{m}^{T}\mathcal{M}^{-1}\textbf{m}=\textbf{m}_{i}^{T}(\mathcal{M}^{-1})^{ij}\textbf{m}_{j}=e^{\phi}\left[\left(m-n\chi\right)^{2}+n^{2}e^{-2\phi}\right],
        \end{equation*}
       \begin{equation}
        \textbf{m}^{T}\textbf{B}=\textbf{m}_{i}\textbf{B}^{i}=mB^{(2)}_{MN}+nC^{(2)}_{MN}
        \label{matrixrelations}
    \end{equation} Therefore substitution of equation (\ref{matrixrelations}) reduces the action as
    \begin{equation}
    \begin{split}
        S_{m,n}=&-T_{D1}\sqrt{\left(m-n\chi\right)^{2}+n^{2}e^{-2\phi}}\int d\tau d\sigma \sqrt{-detG_{MN}\partial_{\alpha}x^{M}\partial_{\beta}x^{N}}\\&+T_{D1}m\epsilon^{\alpha\beta}\int d\tau d\sigma B^{(2)}_{MN}\partial_{\alpha}x^{M}\partial_{\beta}x^{N}+T_{D1}n\epsilon^{\alpha\beta}\int d\tau d\sigma C^{(2)}_{MN}\partial_{\alpha}x^{M}\partial_{\beta}x^{N}
        \end{split}
        \label{NG action}
    \end{equation}It is to be noted that this action is taken in string frame with metric $G_{MN}$ given by $G^{string}_{MN}=e^{\frac{\phi}{2}}g^{E}_{MN}$. The nonlinearity of this action due to the presence of the square root of the determinant results in the equations of motion which are not desirable. To avoid such discrepancy it may be modified by introducing an auxiliary metric $h_{\alpha\beta}$. With such consideration, the action becomes 
    \begin{equation}
    \begin{split}
        S_{m,n}=-\frac{\tau_{(m,n)}}{2}\int d\tau d\sigma \sqrt{-h}h^{\alpha\beta}G_{\alpha\beta}&+q_{(m,n)}\epsilon^{\alpha\beta}\int d\tau d\sigma B^{(2)}_{MN}\partial_{\alpha}x^{M}\partial_{\beta}x^{N}\\&+\tilde{q}_{(m,n)}\epsilon^{\alpha\beta}\int d\tau d\sigma C^{(2)}_{MN}\partial_{\alpha}x^{M}\partial_{\beta}x^{N}
     \end{split}
        \label{Polyakov}
    \end{equation}where, 
    \begin{equation}
        \tau_{(m,n)}=T_{D1}\sqrt{\left(m-n\chi\right)^{2}+n^{2}e^{-2\phi}},~q_{(m,n)}=m\frac{T_{F}}{g_{s}},~\tilde{q}_{(m,n)}=nT_{D1}
        \label{tensionsI}
    \end{equation}Tensions of D1-string and fundamental string, respectively, are
    \begin{equation}
        T_{D1}=\frac{1}{2\pi\alpha^{'}g_{s}},~~T_{F}=\frac{1}{2\pi\alpha^{'}}
        \label{tensionII}
    \end{equation}with $g_{s}$ being the string coupling.   It is evident that there is a factor $\frac{1}{2}$ missing in the flux terms in the action unlike the action for fundamental string as it is derived from the DBI action of D1 brane instead of the Polykov action for fundamental string considered in \cite{Hoare:2013pma, Hoare:2013lja}. From this action equation of motion for $h_{\alpha\beta}$ may be obtained as
     \begin{equation}
     T_{\alpha\beta}=-\frac{2}{\sqrt{-h}}\frac{\partial S_{m,n}}{\partial h_{\alpha\beta}}=\frac{1}{2}h_{\alpha\beta}h^{\rho\xi}G_{\rho\xi}-G_{\alpha\beta}=0
    \end{equation}which gives nothing but tracelessness of the stress-energy tensor and can reproduce the action (\ref{NG action}) from (\ref{Polyakov}) by inserting $h_{\alpha\beta}$=$G_{\alpha\beta}$. Equation of motion for $x^{M}$ can be easily found as
    \begin{equation}
        \begin{split}
            &-2\partial_{\alpha}\left[\sqrt{-h}h^{\alpha\beta}G_{MN}\partial_{\beta}x^{N}\right]+\sqrt{-h}h^{\alpha\beta}\partial_{\alpha}x^{K}\partial_{\beta}x^{L}\partial_{M}G_{KL}\\&+2q_{(m,n)}H_{MKL}\partial_{\tau}x^{K}\partial_{\sigma}x^{L}+2\tilde{q}_{(m,n)}F_{MKL}\partial_{\tau}x^{K}\partial_{\sigma}x^{L}=0
        \end{split}
    \end{equation}where,
    \begin{equation}  
 H_{MNK}=\partial_{M}B^{(2)}_{NK}+\partial_{N}B^{(2)}_{KM}+\partial_{K}B^{(2)}_{MN},~F_{MNK}=\partial_{M}C^{(2)}_{NK}+\partial_{N}C^{(2)}_{KM}+\partial_{K}C^{(2)}_{MN}
    \end{equation} 
In what follows, we will use the metric for $AdS_{3}\times S^{3}$ background in terms of the global background coordinates, 
    \begin{equation}
        ds^{2}=L^{2}\left[-\cosh^{2}\rho dt^{2}+d\rho^{2}+\sinh^{2}\rho d\phi^{2}+d\theta^{2}+\sin^{2}\theta d\phi_{1}^{2}+\cos^{2}\theta d\phi_{2}^{2}\right] \ .
    \end{equation}The accompanying flux components are given by
    \begin{equation}
    \begin{split}
        &B^{(2)}_{t\phi}=L^{2}q^{2}\sinh^{2}\rho,~~B^{(2)}_{\phi_{1}\phi_{2}}=-L^{2}q^{2}\cos^{2}\theta\\&
        C^{(2)}_{t\phi}=L^{2}\sqrt{1-q^{2}}\sinh^{2}\rho,~~C^{(2)}_{\phi_{1}\phi_{2}}=-L^{2}\sqrt{1-q^{2}}\cos^{2}\theta
        \end{split}
    \end{equation}where $q$ and $\sqrt{1-q^{2}}=\tilde{q}$, say, are the parameters associated to field strengths of NS-NS and R-R fluxes respectively. It satisfies, $0\leq q\leq 1$ and $q^{2}+\tilde{q}^{2}=1$. For $q=0$, it is a case of pure RR flux and the worldsheet theory can be described in terms of a Green-Schwarz coset \cite{Metsaev:1998it,Babichenko:2009dk}. On the other hand, for $q=1$ it corresponds to pure NS-NS background, where the theory can be described in terms of a class of supersymmetric WZW model. For intermediate value of $q$ the theory has not been completely understood till date. 
    \subsection{Lagrangian and Hamiltonian formulation}
 To specify the geometry of $AdS_{3}$ and $S_{3}$ it is convenient to consider the embedding coordinates to be $Y_{i}$'s and $W_{i}$'s respectively. These are related to the global coordinates as 
    \begin{equation}
        Y_{1}+iY_{2}=\sinh{\rho}e^{i\phi},~~Y_{3}+iY_{0}=\cosh{\rho}e^{it}
        \label{Y}
    \end{equation}
    \begin{equation}
        W_{1}+iW_{2}=\sin{\theta}e^{i\phi_{1}},~~ W_{3}+iW_{4}=\cos{\theta}e^{i\phi_{2}}
        \label{W}
    \end{equation}The relations (\ref{Y}) and (\ref{W}) satisfy the following constraints \cite{Arutyunov:2003uj, Arutyunov:2003za}
    \begin{equation}
        -Y_{0}^{2}+Y_{1}^{2}+Y_{2}^{2}-Y_{3}^{2}=-1
        \label{embedding1}
    \end{equation}
    \begin{equation} 
        W_{1}^{2}+W_{2}^{2}+W_{3}^{2}+W_{4}^{2}=1
        \label{embedding2}
    \end{equation}which precisely define the geometries of $AdS_{3}$ and $S_{3}$ respectively. As we are interested in the dynamics in $R_t \times S^3$, we take $Y_{1}=Y_{2}=0$ so that $Y_{3}+iY_{0}=e^{it}$. Taking $r_{1}(\xi)=\sin{\theta},r_{2}(\xi)=\cos{\theta},\Phi_{1}(\xi)=\phi_{1}$ and $\Phi_{2}(\xi)=\phi_{2}$ with $\xi=\alpha\sigma+\beta\tau$, the Lagrangian of $(m,n)$ string in $R_t\times S^{3}$ geometry with mixed three form fluxes becomes
    \begin{equation}
    \begin{split}
        &\mathcal{L}=-\frac{\tau_{(m,n)}L^{2}}{2}\left[(\partial_{\tau}t)^{2}-(\partial_{\sigma}t)^{2}+(\partial_{\sigma}r_{1})^{2}-(\partial_{\tau}r_{1})^{2}+(\partial_{\sigma}r_{2})^{2}-(\partial_{\tau}r_{2})^{2}\right]\\&-\frac{\tau_{(m,n)}L^{2}}{2}\left[r_{1}^{2}\big\{(\partial_{\sigma}\Phi_{1})^{2}-(\partial_{\tau}\Phi_{1})^{2}\big\}+r_{2}^{2}\big\{(\partial_{\sigma}\Phi_{2})^{2}-(\partial_{\tau}\Phi_{2})^{2}\big\}\right]+\frac{\Lambda}{2}\left(r_{1}^{2}+r_{2}^{2}-1\right)\\&-q_{(m,n)}qL^{2}\cos^{2}{\theta}\left[(\partial_{\sigma}\Phi_{1})(\partial_{\tau}\Phi_{2})-(\partial_{\tau}\Phi_{1})(\partial_{\sigma}\Phi_{2})\right]\\&-\tilde{q}_{(m,n)}L^{2}\sqrt{1-q^{2}}\cos^{2}\theta\left[(\partial_{\sigma}\Phi_{1})(\partial_{\tau}\Phi_{2})-(\partial_{\tau}\Phi_{1})(\partial_{\sigma}\Phi_{2})\right]
    \end{split}
    \label{targetspaceLagrangian}
    \end{equation}where $\Lambda$ is the Lagrange multiplier. The same reason as explained in the case of the action (\ref{Polyakov}) supports for the missing factor $\frac{1}{2}$ in the flux terms in equation (\ref{targetspaceLagrangian}) unlike the Lagrangian obtained in \cite{Hoare:2013pma} where fundamental string is taken as a probe. Let us consider the following parametrization
    \begin{equation}
t=\kappa\tau,~~r_{a}=r_{a}(\xi),~~\Phi_a=\Phi_{a}(\xi)=\omega_{a}\tau+f_{a}(\xi),~~\xi=\alpha\sigma+\beta\tau.
\label{ansatz}
    \end{equation}where $\kappa,\omega_{a},\alpha$ and $\beta$ are constants. For closed strings $r_{a}$ and $f_{a}$ should follow periodic conditions 
    \begin{equation}
        r_{a}(\xi+2\pi\alpha)=r(\xi),~~f_{a}(\xi+2\pi\alpha)=f_{a}(\xi)+2\pi\bar{m}_{a},
    \end{equation}$\bar{m}_{a}$ being integer winding numbers. Using such parametrization  for the string rotating in $R_{t}\times S^{3}$, the Lagrangian becomes,
    \begin{equation}
    \begin{split}
        \mathcal{L}=&-\frac{\tau_{(m,n)}L^{2}}{2}\left[\kappa^{2}+(\alpha^{2}-\beta^{2})
 \sum_{a=1}^{2}\left({r_{a}}^{'2}+{r_{a}}^{2}\left(f_{a}^{'}-
 \frac{\beta\omega_{a}}{\alpha^{2}-\beta^{2}}\right)^{2}-\frac{\alpha^{2}
 \omega_{a}^{2}r_{a}^{2}}{(\alpha^{2}-\beta^{2})^{2}}\right)\right]\\&-\left(q_{(m,n)}q+\tilde{q}_{(m,n)}\sqrt{1-q^{2}}\right)L^{2}\alpha^{2}r_{2}^{2}\left(\omega_{2}f_{1}^{'}-\omega_{1}f_{2}^{'}\right)+\frac{\Lambda}{2}\left(r_{1}^{2}+r_{2}^{2}-1\right) \ ,
 \end{split}
 \label{lagrangian 1}
 \end{equation} where derivatives with respect to $\xi$ are denoted by prime. Equations of motion for $f_{a}$'s can be calculated as  
 \begin{equation}
     f_{a}^{'}=\frac{1}{\alpha^{2}-\beta^{2}}\left[\frac{C_{a}}{r_{a}^{2}}+\beta \omega_{a}+\frac{Q\alpha^{2}r_{2}^{2}\omega_{b}}{r_{a}^{2}}\epsilon_{ba}\right]
     \label{equationf}
    \end{equation}where $\epsilon_{ba}=1$ and $C_{a}$'s are suitable integration constants. Also 
    \begin{equation}
    Q=\frac{q_{(m,n)}q+\tilde{q}_{(m,n)}\sqrt{1-q^{2}}}{\tau_{(m,n)}}=\frac{mq+n\sqrt{1-q^{2}}}{\sqrt{(m-n\chi)^{2}+n^{2}e^{-2\phi}}}
    \label{fluxexpression}
 \end{equation}where we have used (\ref{tensionsI}) and (\ref{tensionII}). Therefore, the flux parameters are involved here in the quantity denoted as $Q$ which eventually contains both the integers $m$ and $n$ along with the axion scalar and dilaton. The special case with $n=0$ obtains from (\ref{fluxexpression}) the flux parameter for probe fundamental string. Putting this expression of $f_{a}^{'}$ in the Lagrangian (\ref{lagrangian 1}) we achieve,
 \begin{equation}
 \begin{split}
     \mathcal{L}=&-\frac{\tau_{(m,n)}L^{2}}{2}\left[\kappa^{2}+\left(\alpha^{2}-\beta^{2}\right)\sum_{a=1}^{2}r_{a}^{'2}+\frac{1}{\left(\alpha^{2}-\beta^{2}\right)}\sum_{a=1}^{2}\left(\frac{C_{a}^{2}+Q^{2}\alpha^{4}r_{2}^{4}\omega_{b}^{2}}{r_{a}^{2}}\right)\right]\\&+\frac{\tau_{(m,n)}L^{2}}{2}\left[\frac{\alpha^{2}}{\alpha^{2}-\beta^{2}}\sum_{a=1}^{2}\left(\omega_{a}^{2}r_{a}^{2}+2C_{a}Q\omega_{b}r_{2}^{2}\epsilon_{ba}\right)\right]+\frac{\Lambda}{2}\left(r_{1}^{2}+r_{2}^{2}-1\right)\\&
     -Q\tau_{(m,n)}\frac{\alpha^{2}L^{2}r_{2}^{2}}{\alpha^{2}-\beta^{2}}\left[\frac{\omega_{2}C_{1}}{r_{1}^{2}}-\frac{C_{2}\omega_{1}}{r_{2}^{2}}+\frac{Q\alpha^{2}(\omega_{1}^{2}r_{1}^{2}+\omega_{2}^{2}r_{2}^{2})}{r_{1}^{2}}\right] \ .
     \end{split}
 \end{equation} The Lagrangian contains both $r_{a}^{2}$ and $\frac{1}{r_{a}^{2}}$-terms representing harmonic oscillator type potential and centrifugal potential barrier respectively. Hence it can be identified as the Lagrangian of one dimensional integrable Neumann-Rosochatius system with extra terms due to presence of mixed flux. We can get the equations of motion for $r_{1}$ and $r_{2}$ as 
 \begin{equation}
     r_{1}^{''}=r_{1}\left[\left(f_{1}^{'}-\frac{\beta\omega_{1}}{\alpha^{2}-\beta^{2}}\right)^{2}-\frac{\alpha^{2}\omega_{1}^{2}}{(\alpha^{2}-\beta^{2})^{2}}\right]+Ar_{1} \ ,
     \label{equationr1}
 \end{equation}and
        \begin{equation}
            r_{2}^{''}=r_{2}\left[\left(f_{2}^{'}-\frac{\beta\omega_{2}}{\alpha^{2}-\beta^{2}}\right)^{2}-\frac{\alpha^{2}\omega_{2}^{2}}{(\alpha^{2}-\beta^{2})^{2}}\right]+Ar_{2}+\frac{2Q\alpha^{2}r_{2}}{(\alpha^{2}-\beta^{2})^{2}}\left(\omega_{2}f_{1}^{'}-\omega_{1}f_{2}^{'}\right) \ ,
            \label{equationr2}
        \end{equation}where $A=\frac{\Lambda}{\tau_{(m,n)}L^{2}}$ is a constant. Putting the expressions of $f_{a}$'s from (\ref{equationf}) in equations (\ref{equationr1}) and (\ref{equationr2}) we achieve,
        \begin{equation}
        \begin{split}
            \left(\alpha^{2}-\beta^{2}\right)r_{1}^{''}+&\frac{1}{\left(\alpha^{2}-\beta^{2}\right)}\left(\frac{C_{1}^{2}}{r_{1}^{3}}+\frac{Q^{2}\alpha^{4}r_{2}^{4}\omega_{2}^{2}}{r_{1}^{3}}\right)+\frac{\alpha^{2}}{\left(\alpha^{2}-\beta^{2}\right)}\omega_{1}^{2}r_{1}-Ar_{1}\\&+\frac{2Q\alpha^{2}r_{2}^{2}}{\left(\alpha^{2}-\beta^{2}\right)}\left[\frac{\omega_{2}C_{1}}{r_{1}^{3}}+\frac{Q\alpha^{2}r_{2}^{2}\omega_{2}^{2}}{r_{1}^{3}}\right]=0 \ ,
            \end{split}
        \end{equation}and 
        \begin{equation}
        \begin{split}
            &\left(\alpha^{2}-\beta^{2}\right)r_{2}^{''}-\frac{2Q^{2}\alpha^{4}r_{2}^{3}\omega_{2}^{2}}{r_{1}^{2}\left(\alpha^{2}-\beta^{2}\right)}-\frac{Q^{2}\alpha^{4}r_{2}\omega_{1}^{2}}{\left(\alpha^{2}-\beta^{2}\right)}+\frac{1}{\left(\alpha^{2}-\beta^{2}\right)}\frac{C_{2}^{2}}{r_{2}^{3}}+\frac{\alpha^{2}}{\left(\alpha^{2}-\beta^{2}\right)}\omega_{2}^{2}r_{2}\\& +\frac{2Q\alpha^{2}}{\left(\alpha^{2}-\beta^{2}\right)}\left(\omega_{2}C_{1}-\omega_{1}C_{2}\right)r_{2}-Ar_{2}-\frac{2Q\alpha^{2}}{\left(\alpha^{2}-\beta^{2}\right)}\left[\frac{\omega_{2}C_{1}r_{2}}{r_{1}^{2}}+\frac{2Q\alpha^{2}r_{2}^{3}\omega_{2}^{2}}{r_{1}^{2}}+Q\alpha^{2}\omega_{1}^{2}r_{2}\right]=0 \ .
            \end{split}
        \end{equation}Above two equations can also be obtained from the Euler-Lagrange equations of the following equivalent Lagrangian of an integrable NR system
        \begin{equation}
             \begin{split}
     \mathcal{L}=&\frac{\left(\alpha^{2}-\beta^{2}\right)}{2}\sum_{a=1}^{2}r_{a}^{'2}-\frac{1}{2\left(\alpha^{2}-\beta^{2}\right)}\sum_{a=1}^{2}\left(\frac{C_{a}^{2}+Q^{2}\alpha^{4}r_{2}^{4}\omega_{b}^{2}}{r_{a}^{2}}\right)\\&+\left[\frac{\alpha^{2}}{2\left(\alpha^{2}-\beta^{2}\right)}\sum_{a=1}^{2}\left(\omega_{a}^{2}r_{a}^{2}+2C_{a}Q\omega_{b}r_{2}^{2}\epsilon_{ba}\right)\right]+\frac{A}{2}\left(r_{1}^{2}+r_{2}^{2}\right)\\&
     -\frac{Q\alpha^{2}r_{2}^{2}}{2\left(\alpha^{2}-\beta^{2}\right)}\left[\frac{\omega_{2}C_{1}}{r_{1}^{2}}-\frac{C_{2}\omega_{1}}{r_{2}^{2}}+\frac{Q\alpha^{2}(\omega_{1}^{2}r_{1}^{2}+\omega_{2}^{2}r_{2}^{2})}{r_{1}^{2}}\right] \, 
     \end{split}
        \end{equation}for general coordinates $r_{1}$ and $r_{2}$. The Virasoro constraints may be expressed as 
        \begin{subequations}
        \begin{align}
            T_{\tau\tau}+T_{\sigma\sigma}=0, ~~~~~~ 
            T_{\sigma\tau}=T_{\tau\sigma}=0 \ .
            \end{align}
        \end{subequations} With the embeddings (\ref{embedding1}) and (\ref{embedding2}) and using the ansatz (\ref{ansatz}), the constraints reduce to
        \begin{equation}
            (\alpha^{2}+\beta^{2})(r_{1}^{'2}+r_{2}^{'2})+r_{1}^{2}\left[\alpha^{2}f_{1}^{'2}+(\omega_{1}+\beta f_{1}^{'})^{2}\right]+r_{2}^{2}\left[\alpha^{2}f_{2}^{'2}+(\omega_{2}+\beta f_{2}^{'})^{2}\right]=\kappa^{2} \, 
            \label{vc1}
        \end{equation}and
        \begin{equation}
\alpha\beta(r_{1}^{'2}+r_{2}^{'2})+\alpha\left[\omega_{1}r_{1}^{2}f_{1}^{'}+\omega_{2}r_{2}^{2}f_{2}^{'}\right]+\alpha\beta\left[f_{1}^{'2}r_{1}^{2}+f_{2}^{'2}r_{2}^{2}\right]=0
\label{vc2}
        \end{equation}respectively. The Hamiltonian of such a system can be written as
        \begin{equation}
                    \begin{split}
     &\mathcal{H}=\frac{\left(\alpha^{2}-\beta^{2}\right)}{2}\sum_{a=1}^{2}r_{a}^{'2}+\frac{1}{2\left(\alpha^{2}-\beta^{2}\right)}\sum_{a=1}^{2}\left(\frac{C_{a}^{2}+Q^{2}\alpha^{4}r_{2}^{4}\omega_{b}^{2}}{r_{a}^{2}}\right)\\&-\left[\frac{\alpha^{2}}{2\left(\alpha^{2}-\beta^{2}\right)}\sum_{a=1}^{2}\left(\omega_{a}^{2}r_{a}^{2}+2C_{a}Q\omega_{b}r_{2}^{2}\epsilon_{ba}\right)\right]+\\&\frac{Q\alpha^{2}r_{2}^{2}}{2\left(\alpha^{2}-\beta^{2}\right)}\left[\frac{\omega_{2}C_{1}}{r_{1}^{2}}-\frac{C_{2}\omega_{1}}{r_{2}^{2}}+\frac{Q\alpha^{2}(\omega_{1}^{2}r_{1}^{2}+\omega_{2}^{2}r_{2}^{2})}{r_{1}^{2}}\right]
     \end{split}
        \end{equation}This again takes the form of the Hamiltonian of integrable one dimensional NR model. Hence it is obvious that classically integrable NR model may represent a nice tool to study for probe $(m,n)$ string in the desired background, even in the presence of flux. 
\subsection{Integrals of Motion}
Any classically integrable system contains infinite number of conserved quantities, also known as integrals of motion, in involution, i.e., they must Poisson commute each other. The integrability of Neumann-Rosochatius system demands the existence of a set of integrals of motion, named as Uhlenbeck constants, in involution. These were first introduced by K. Uhlenbeck \cite{Uhlenbeck}. In the case of closed rotating string in $S^{3}$ there are two integrals of motion $I_{1}$ and $I_{2}$ constrained by the relation $I_{1}+I_{2}=1$. A general form of Uhlenbeck integrals of motion for the case of closed rotating string can be written as
        \begin{equation}
            I_{a}=\alpha^{2}x_{a}x_{a}^{'}+\sum_{b\neq a}\frac{|\bar{x_{b}}p_{a}-x_{a}\bar{p_{b}}|^{2}}{\omega_{a}^{2}-\omega_{b}^{2}} \ ,
        \end{equation}which eventually gives
        \begin{equation}
            I_{a}=\alpha^{2}r_{a}^{2}+\left(\alpha^{2}-\beta^{2}\right)^{2}\sum_{b \neq a}\frac{\left(r_{a}^{'}r_{b}-r_{a}r_{b}^{'}\right)^{2}}{\omega_{a}^{2}-\omega_{b}^{2}}+\sum_{b\neq a}\frac{1}{\omega_{a}^{2}-\omega_{b}^{2}}\left(\frac{C_{a}^{2}r_{b}^{2}}{r_{a}^{2}}+\frac{C_{b}^{2}r_{a}^{2}}{r_{b}^{2}}\right) \ ,
        \end{equation}for some arbitrary values of constants $C_{a}$'s. Here we have used the embeddings (\ref{embedding1}),\\(\ref{embedding2}) and ansatz (\ref{ansatz}) for closed string rotating on $S^{3}$ with $x_{a}(\xi)=r_{a}e^{f_{a}(\xi)}$ \cite{Kruczenski:2006pk}.
To find out the integrals of motion in the deformed background, let us proceed with the methodology described in \cite{Hernandez:2014eta,Hernandez:2015nba} where a term $g=g(r_{1},r_{2},Q)$ representing the deformation due to  fluxes is added without affecting the integrable properties of $I_a$ such that,
        \begin{equation}
        \begin{split}
            \bar{I}_{a}=\alpha^{2}r_{a}^{2}+\left(\alpha^{2}-\beta^{2}\right)^{2}\sum_{b \neq a}\frac{\left(r_{a}^{'}r_{b}-r_{a}r_{b}^{'}\right)^{2}}{\omega_{a}^{2}-\omega_{b}^{2}}+&\sum_{b\neq a}\frac{1}{\omega_{a}^{2}-\omega_{b}^{2}}\left(\frac{C_{a}^{2}r_{b}^{2}}{r_{a}^{2}}+\frac{C_{b}^{2}r_{a}^{2}}{r_{b}^{2}}\right)\\&+\sum_{b\neq a}\frac{2g}{\omega_{a}^{2}-\omega_{b}^{2}} \ .
            \end{split}
        \end{equation}The function $g$ may be derived by setting $\bar{I_{a}^{'}}=0$. This will yield,
        \begin{equation}
            g=(\omega_{1}^{2}-\omega_{2}^{2})r_{1}^{2}\alpha^{2}-(\alpha^{2}-\beta^{2})(r_{1}^{'2}+r_{2}^{'2})-\left(\frac{C_{1}^{2}}{r_{1}^{2}}+\frac{C_{2}^{2}r_{1}^{2}}{r_{2}^{2}}\right)
        \end{equation}Now, by using the constraints 
        \begin{equation}
            r_{1}^{2}+r_{2}^{2}=1,~~r_{1}r_{1}^{'}+r_{2}r_{2}^{'}=0,~~r_{1}r_{1}^{''}+r_{1}^{'2}+r_{2}^{'2}+r_{2}r_{2}^{''} = 0
        \end{equation}and the equations (\ref{equationr1}), (\ref{equationr2}) and (\ref{equationf}), we get the integral of motion in the following form:
        \begin{equation}
        \begin{split}
\bar{I_{1}}=&\frac{\alpha^{2}-\beta^{2}}{\omega_{1}^{2}-\omega_{2}^{2}}\left(r_{1}r_{2}^{'}-r_{1}^{'}r_{2}\right)^{2}+\frac{2}{\omega_{1}^{2}-\omega_{2}^{2}}\left[\left(\frac{C_{1}-Q\alpha^{2}r_{2}^{2}\omega_{2}}{r_{1}}\right)^{2}+\left(\frac{C_{2}+Q\alpha^{2}r_{2}^{2}\omega_{1}}{r_{2}}\right)^{2}\right]\\&-\frac{2\alpha^2}{\omega_{1}^{2}-\omega_{2}^{2}}\left[\left(1+\frac{2Q^{2}\alpha^{2}r_{2}^{2}}{r_{1}^{2}}\right)\left(\omega_{1}^{2}r_{1}^{2}+\omega_{2}^{2}r_{2}^{2}\right)+2Qr_{2}^{2}\left(\frac{C_{1}\omega_{2}}{r_{1}^{2}}-\frac{C_{2}\omega_{1}}{r_{2}^{2}}\right)\right]\\&+\frac{1}{\omega_{1}^{2}-\omega_{2}^{2}}\left(\frac{C_{1}^{2}}{r_{1}^{2}}+\frac{C_{2}^{2}r_{1}^{2}}{r_{2}^{2}}\right) \ .
\label{integralofmotion}
            \end{split}
        \end{equation}
It is henceforth clear that in the absence of flux, the deformed constants satisfy the relation $\sum_{a=1}^{2}\bar{I}_{a}=1$.
    \subsection{Elliptic solutions for $r_{1}$ and $r_{2}$}
In this section we present the solutions for such system of $(m,n)$ string rotating in $S^{3}$ with two different angular momenta. In this context we can introduce three parameters on a sphere like $\zeta_{1},\zeta_{2}$ and $\zeta_{3}$ to express the roots of the equation in ellipsoidal coordinate $\zeta$  
        \begin{equation}
            \frac{r_{1}^{2}}{\zeta-\omega_{1}^{2}}+\frac{r_{2}^{2}}{\zeta-\omega_{2}^{2}}=0 \ .
            \label{ellipsoidal}
        \end{equation}With angular frequencies $\omega_{1}<\omega_{2}$, this equation is defined on a sphere while being invariant under $r_{a}\rightarrow \lambda r_{a}$ \cite{Babelon:1992rb}. The range of ellipsoidal coordinate is $\omega_{1}^{2}\leq \zeta \leq \omega_{2}^{2}$ for which $\zeta$ covers $\frac{1}{4}$th of a sphere corresponding to $r_{i}\geq 0$ and the whole sphere may be thought of as a covering of the domain of $\zeta$ having branches along its boundary. If we enter the expressions of $r_1^2$ and $r_2^2$ in terms of $\zeta$ into the equations of motion we may get a set of second order differential equations for $\zeta$. But for the sake of simplicity, we inject this ellipsoidal coordinate $\zeta$ into the Uhlenbeck integrals of motion \cite{Arutyunov:2003uj, Arutyunov:2003za} and can have a first order differential equation for $\zeta$. Using the constraint $r_1^2+r_2^2=1$ corresponding to spherical geometry in  equation (\ref{ellipsoidal}) we find
        \begin{equation}
            r_{1}^{2}=\frac{\omega_{1}^{2}-\zeta}{\omega_{1}^{2}-\omega_{2}^{2}},~~r_{2}^{2}=\frac{\zeta-\omega_{2}^{2}}{\omega_{1}^{2}-\omega_{2}^{2}},~~\left(r_{1}r_{2}^{'}-r_{1}^{'}r_{2}\right)^{2}=\frac{\zeta^{'2}}{4(\omega_{1}^{2}-\zeta)(\zeta-\omega_{2}^{2})} \ .
        \end{equation} The integral of motion (\ref{integralofmotion}) in terms of the ellipsoidal coordinate takes the following form
        \begin{equation}
            \begin{split}
                \bar{I}_{1}=&\frac{\alpha^{2}-\beta^{2}}{\omega^{2}_{1}-\omega_{2}^{2}}\frac{\zeta^{'2}}{4(\omega_{1}^{2}-\zeta)(\zeta-\omega_{2}^{2})}+\frac{2Q^{2}\alpha^{4}(\zeta-\omega_{2}^{2})(\zeta+\omega_{1}^{2}+\omega_{2}^{2})}{(\omega_{1}^{2}-\omega_{2}^{2})(\omega_{1}^{2}-\zeta)}+\\&\frac{2\alpha^{2}(\omega_{1}^{2}-\zeta)}{(\omega_{1}^{2}-\omega_{2}^{2})}+\frac{2\alpha^{2}\omega_{2}^{2}Q}{(\omega_{1}^{2}-\zeta)}+\frac{2\alpha^{2}}{\omega_{1}^{2}-\omega_{2}^{2}}\left[Q\omega_{1}^{2}+\omega_{2}^{2}(1-Q)\right] \ .
            \end{split}
        \end{equation}
Solving for $\zeta^{'2}$ from this expression  of deformed Uhlenbeck constant we explore that 
        \begin{equation}
            \zeta^{'2}=-4P_{3}(\zeta),~~ P_{3}(\zeta)=\frac{2\alpha^{2}(1-Q)}{\alpha^{2}-\beta^{2}}\prod_{i=1}^{3}\left(\zeta-\zeta_{i}\right) \ .
            \label{elliptic}
        \end{equation}
Here $P_{3}(\zeta)$ is evidently a third order polynomial  defining an elliptic curve\\ $s^{2}+P_{3}(\zeta)=0$, $s_i$ and $\zeta_i$ being the general coordinates on that elliptic curve. Now changing the variables as 
        \begin{equation} 
            \zeta=\zeta_{2}+(\zeta_{3}-\zeta_{2})\eta^{2}
            \label{variablechange}
        \end{equation}and substituting for $\eta$ in equation (\ref{elliptic}) we get 
        \begin{equation}
            \eta(\xi)=\hbox{cn}\left(\alpha\xi\sqrt{\frac{2\left(1-Q\right)\left(\zeta_{3}-\zeta_{1}\right)}{\alpha^{2}-\beta^{2}}}+\xi_{0}, k\right) \ ,
        \end{equation}where $k=\frac{\zeta_{3}-\zeta_{2}}{\zeta_{3}-\zeta_{1}}$ is the elliptic modulus and $\xi_{0}$ is the integration constant which may be set to zero by a rotation. This gives the expression for $r_{1}$ from equation (\ref{ellipsoidal}) as
        \begin{equation}
            r_{1}^{2}(\xi)=\frac{\zeta_{3}-\omega_{1}^{2}}{\omega_{2}^{2}-\omega_{1}^{2}}+\frac{\zeta_{2}-\zeta_{3}}{\omega_{2}^{2}-\omega_{1}^{2}}\hbox{sn}^{2}\left(\alpha\xi\sqrt{\frac{2\left(1-Q\right)\left(\zeta_{3}-\zeta_{1}\right)}{\alpha^{2}-\beta^{2}}},k\right)
            \label{solutionr1}
        \end{equation} and similarly the expression for $r_{2}$ is 
        \begin{equation}
            r_{2}^{2}(\xi)=\frac{\omega_{2}^{2}-\zeta_{3}}{\omega_{2}^{2}-\omega_{1}^{2}}-\frac{\zeta_{2}-\zeta_{3}}{\omega_{2}^{2}-\omega_{1}^{2}}\hbox{sn}^{2}\left(\alpha\xi\sqrt{\frac{2\left(1-Q\right)\left(\zeta_{3}-\zeta_{1}\right)}{\alpha^{2}-\beta^{2}}},k\right)
            \label{solutionr2}
        \end{equation}It is obvious from these expressions that the constraint $r_{1}^{2}+r_{2}^{2}=1$ is satisfied. The range of elliptic modulus $k$ should be $0<k<1$ which eventually needs $\zeta_{1}<\zeta_{2}<\zeta_{3}$. Also we can achieve circular type solution for $\omega_{1}^{2}\leq\zeta_{2,3}\leq\omega_{2}^{2}$ in the codomain of equations (\ref{solutionr1}) and (\ref{solutionr2}) between $0$ and $1$ provided that no such restriction is needed for $\zeta_{1}$.
        \subsection{Conserved charges and dispersion relation}
In this section we are interested in finding out the scaling relation among various conserved charges. The conserved energy and angular momenta for this system are expressed as
\begin{equation}
    E=-\int{d\sigma \frac{\partial\mathcal{L}}{\partial(\partial_{\tau}t)}},~~ J_{a}=\int{d \sigma\frac{\partial\mathcal{L}}{\partial(\partial_{\tau}\Phi_{a})}},
\end{equation} with $a=1,2$ 
Therefore using the Lagrangian (\ref{targetspaceLagrangian}), conserved quantities may be expressed as
\begin{subequations}
   \begin{align}
       &E=2\pi\tau_{(m,n)}L^{2}\kappa \ , \\&
 J_{1}=\frac{1}{\alpha}\int_{0}^{2\pi}d\xi\tau_{(m,n)}L^{2}\left[r_{1}^{2}\left(\omega_{1}^{2}+\beta f_{1}^{'}\right)+Q\alpha r_{2}^{2}f_{2}^{'}\right]\label{current1} \ , \\&
   J_{2}=\frac{1}{\alpha}\int_{0}^{2\pi}d\xi\tau_{(m,n)}L^{2}\left[r_{2}^{2}\left(\omega_{2}^{2}+\beta f_{2}^{'}\right)+Q\alpha r_{2}^{2}f_{1}^{'}\right] \ . \label{current2}
   \end{align}
   \end{subequations}
\subsection{Constant radii solutions}
A convenient method of deriving the energy-angular momenta dispersion relation is to employ the equations (\ref{solutionr1}) and (\ref{solutionr2}) in (\ref{current1}) and (\ref{current2}) and to write the energy in terms of winding numbers $\bar{m}_i$'s, angular momenta $J$'s and Uhlenbeck constants $\bar{I}_{i}$'s. However, there are some important limits of the parameters $\zeta_i$'s for which the computation of dispersion relation can be simplified. The corresponding choices of the parameters may be considered such that the discriminant of $P_3(\zeta)$ reduces to zero. The possible choices are : (i) constant radii case which is achieved with the limit $\zeta_2\rightarrow \zeta_3$, (ii) $\zeta_1=\zeta_2$ which gives $k=1$ and (iii) $\zeta_1=\zeta_2=\zeta_3$ which can only be solved for equal angular frequencies $\omega_1=\omega_2$. We are interested in deriving the constant radii solutions for the conserved energy and momenta which can be obtained by taking the limit $\zeta_{2}\rightarrow \zeta_{3}$ in the equations (\ref{solutionr1}) and (\ref{solutionr2}). These type of solutions have been constructed by using Neumann-Rosochatius integrable system in \cite{Hernandez:2014eta} for circular strings without incorporating the Weierstrass elliptic functions. With this limit, the radial coordinates become constant such as $r_{1}=\sqrt{\frac{\zeta_{3}-\omega_{1}^{2}}{\omega_{2}^{2}-\omega_{1}^{2}}}=a_{1}$, say and $r_{2}=\sqrt{\frac{\omega_{2}^{2}-\zeta_{3}}{\omega_{2}^{2}-\omega_{1}^{2}}}=a_{2}$, say. Then the derivatives of the angles become constant, thereby giving 
        \begin{equation}
            f_{a}=\bar{m}_{a}\xi+f_{0a} \ ,
        \end{equation}where the integration constants $f_{0a}$ can be chosen to be zero through a rotation and $\bar{m}_{a}=\frac{1}{\alpha^{2}-\beta^{2}}\left[\frac{C_{a}}{a_{a}^{2}}+\beta \omega_{a}+\frac{Q\alpha^{2}a_{2}^{2}\omega_{b}}{a_{a}^{2}}\epsilon_{ba}\right]$ are assumed to be the constant integer windings of the string satisfying the closed string periodicity conditions. Also the Virasoro constraints (\ref{vc1}), (\ref{vc2}) and currents (\ref{current1}), (\ref{current2}) yield
        \begin{equation}
            a_{1}^{2}=\frac{\bar{m}_{2}(\omega_{2}+\beta\bar{m}_{2})}{\bar{m}_{2}(\omega_{2}+\beta\bar{m}_{2})-\bar{m}_{1}(\omega_{1}+\beta \bar{m}_{1})},~~ a_{2}^{2}=\frac{\bar{m}_{1}(\omega_{1}+\beta\bar{m}_{1})}{\bar{m}_{1}(\omega_{1}+\beta\bar{m}_{1})-\bar{m}_{2}(\omega_{2}+\beta \bar{m}_{2})} \ ,
            \label{radius}
        \end{equation}and 
        \begin{equation}
            \bar{m}_{1}J_{1}+\bar{m}_{2}J_{2}=0 \ .
            \label{mJ eq}
        \end{equation} The above relations along with the conserved quantities yield from (\ref{vc1}),
        \begin{equation}
        \begin{split}
            \frac{E^{2}}{\alpha^{2}}=\left(J_{1}+J_{2}\right)^{2}+\frac{(1-w)^{2}}{w}J_{1}J_{2}&-2QT_{(m,n)}\bar{m}_{1}\left(J_{2}+J_{1}w\right)\\&+T_{(m,n)}^{2}\left(\bar{m}_{1}\bar{m}_{2}-Q^{2}\bar{m}_{1}^{2}w\right)\left(\frac{\bar{m}_{1}-\bar{m}_{2}w}{\bar{m}_{2}-\bar{m}_{1}w}\right) \ ,
            \end{split}
            \label{energy}
        \end{equation}where we have used
        $T_{(m,n)}=2\pi \tau_{(m,n)}L^{2}$ and $w=\frac{\omega_{1}+\beta\bar{m}_{1}}{\omega_{2}+\beta\bar{m}_{2}}$. To derive the dispersion relation it would be convenient to find out $w$ in terms of windings $\bar{m}_{a}$ and angular momenta $J_{a}$. To do this let us write the reduced  equations of motion for constant radii solutions as
        \begin{equation}
            \bar{m}_{1}^{2}-\frac{2\beta\omega_{1}\bar{m}_{1}}{\alpha^{2}-\beta^{2}}-\frac{\omega_{1}^{2}}{\alpha^{2}-\beta^{2}}+A=0 \ ,
            \label{EOM3}
        \end{equation}
        \begin{equation}
            \bar{m}_{2}^{2}-\frac{2\beta\omega_{2}\bar{m}_{2}}{\alpha^{2}-\beta^{2}}-\frac{\omega_{2}^{2}}{\alpha^{2}-\beta^{2}}+A+\frac{2Q\alpha^{2}}{(\alpha^{2}-\beta^{2})^{2}}\left(\omega_{2}\bar{m}_{1}-\omega_{1}\bar{m}_{2}\right)=0 \ .
            \label{EOM4}
        \end{equation}Now adding equations (\ref{current1}) and (\ref{current2}), subtracting equation (\ref{EOM4}) from (\ref{EOM3}) and solving the resulting system of equations we get the relation 
        \begin{equation}
            \begin{split}
            &\bar{m}_{1}^{2}-\bar{m}_{2}^{2}+\frac{2\beta}{\alpha^{2}-\beta^{2}}\left[\frac{\bar{m}_{2}}{w}(\omega_{1}+\beta\bar{m}_{1})-\beta\bar{m}_{2}^{2}-\omega_{1}\bar{m}_{1}\right]\\&+\frac{\left[\frac{1}{w}(\omega_{1}+\beta\bar{m}_{1})-\beta\bar{m}_{2}\right]^{2}-\omega_{1}^{2}}{\alpha^{2}-\beta^{2}}-\frac{2Q\alpha^{2}}{(\alpha^{2}-\beta^{2})^{2}}\left[\frac{\bar{m}_{1}}{w}(\omega_{1}+\beta\bar{m}_{1})-\beta\bar{m}_{2}\bar{m}_{1}-\omega_{1}\bar{m}_{2}\right]=0 \ ,
            \end{split}
        \end{equation}where $\omega_{1}$ is expressed as
        \begin{equation}
            \omega_{1}=\frac{\bar{m}_{1}w\left[J-T_{(m,n)}Q\alpha\left(\bar{m}_{1}-\bar{m}_{2}\right)\right]-\bar{m}_{2}J}{T_{(m,n)}(\bar{m}_{1}-\bar{m}_{2})}-\beta \bar{m}_{1} \ ,
        \end{equation} Here $J=J_{1}+J_{2}$ is the total angular momentum. Eliminating $\omega_{1}$ from these two relations we are left with a quartic equation of $w$ given by \begin{equation}
        \begin{split}
            &\left(1-w^{2}\right)\left(\bar{m}_{1}w-\bar{m}_{2}\right)^{2}J^{2}-4QT_{(m,n)}\bar{m}_{1}w\left(\bar{m}_{1}w-\bar{m}_{2}\right)\left(\bar{m}_{1}-\bar{m}_{2}\right)\left(1-w^{2}\right)\\&+\frac{w^{2}Q^{2}T_{(m,n)}^{2}}{J^{2}}\left(\bar{m}_{1}-\bar{m}_{2}\right)\left[4\bar{m}_{1}^{2}\left(1-w^{2}\right)\left(\bar{m}_{1}-\bar{m}_{2}\right)+\bar{m}_{2}\left(\bar{m}_{2}-w\right)\right]\\&+\frac{w^{2}T_{(m,n)}^{2}}{J^{2}}\left(\bar{m}_{1}-\bar{m}_{2}\right)^{3}\left(\bar{m}_{1}+\bar{m}_{2}\right)\left(1-Q^{2}\right)=0 \ .
            \end{split}
            \label{quartic}
        \end{equation}Instead of solving this equation explicitly, it is convenient to get an approximate solution as a power series expansion in large $J$, i.e., small $\frac{T_{(m,n)}}{J}$ which is given by
        \begin{equation}
            w=1+\frac{QT_{(m,n)}}{J}\frac{\bar{m}_{2}}{\bar{m}_{1}}\left(\bar{m}_{1}-\bar{m}_{2}\right)-\frac{Q^{2}T_{(m,n)}^{2}}{J^{2}}\frac{\bar{m}_{1}}{2\bar{m}_{2}^{2}}\left(\bar{m}_{1}-\bar{m}_{2}\right)^{2}+\mathcal{O}\left(\frac{T_{(m,n)}}{J}\right)^{3} \ .
            \label{omega}
        \end{equation} where we have neglected the other higher order terms in the series expansion. Substituting equation (\ref{omega}) in equation (\ref{energy}) we get a relation between energy and angular momenta as follows
\begin{eqnarray}
E^{2} = J^{2}-2QT_{(m,n)}\bar{m}_{1}J+T_{(m,n)}^{2}\frac{\bar{m}_{2}^{2}}{\bar{m}_{1}^{2}}\left(\bar{m}_{1}^{2}-\bar{m}_{2}^{2}\right)\left(1-Q^{2}\right) -\frac{2Q^{2}T_{(m,n)}^{2}}{J}J_{1}\bar{m}_{1}\times \nonumber \\  
\times \left(\bar{m}_{1}-\bar{m}_{2}\right)
\left[\frac{\bar{m}_{2}}{\bar{m}_{1}}-\frac{QT_{(m,n)}}{J}\frac{\bar{m}_{1}}{2\bar{m}_{2}^{2}}\left(\bar{m}_{1}-\bar{m}_{2}\right)\right] \ ,
 \nonumber \\
        \end{eqnarray} where we have considered $\alpha=1$. For two equal angular momentum $J_1 = J_2$ (which sets $\bar{m}_{1}= -\bar{m}_{2}=
        \bar{m}$) , one gets
        \begin{equation}
            E^{2}=J^{2}-4\pi \tau_{(m,n)}L^2 Q \bar{m} J+2\pi^2\tau^{2}_{(m,n)}L^4Q^{2}\bar{m}^{2} \, 
        \end{equation}Again the solution with constant and equal radii demands $\omega=1$ from the expressions (\ref{radius}) and (\ref{mJ eq}). This eventually yields the dispersion relation as
\begin{equation}
E=J-2\pi \tau_{(m,n)}L^2 Q \bar{m} +\frac{\pi^2\tau^{2}_{(m,n)}L^4\left(1-Q^{2}\right)\bar{m}^{2}}{J}
    \end{equation}Here, $T_{(m,n)}=2\pi\tau_{(m,n)}L^2$. The dispersion relation obtained for the equal and constant radii case quite matches with those for rigid circular string rotating in $R\times S^3$ with fundamental string as the natural probe \cite{Hoare:2013lja, Hoare:2013ida} except the flux parameter  which in this case is given as $Q$ described in (\ref{fluxexpression}). For large $J$, the above expression reduces to $E\approx J$. Such linear dependence of $E$ on $J$ stems for the identification of the string state with the gauge theory operator corresponding to the maximal-spin state of XXX spin chain.
     \section{NR system for pulsating (m,n) strings in $AdS_{3}\times S^{3}$ with mixed flux}
     In this section, we study the circular closed string pulsating in $R_{t}\times S^{3}$ by using a deformed integrable one dimensional Neumann-Rosochatius model. We use the following embedding 
     \begin{subequations}
\begin{align}
&Y_{3}+iY_{0}=\cosh{\rho}e^{it}=z_{0}(\tau)e^{ih_{0}(\tau)}\ ,\\&
W_{1}+iW_{2}=\sin{\theta}e^{i\phi_{1}}=r_{1}(\tau)e^{i(f_{1}(\tau)+\bar{m}_{1}\sigma)} \ ,\\& W_{3}+iW_{4}=\cos{\theta}e^{i\phi_{2}}=r_{2}(\tau)e^{i(f_{2}(\tau)+\bar{m}_{2}\sigma)} \ .
    \end{align}
   \end{subequations} These will follow the relations between global and local coordinates as
   \begin{eqnarray}
   &\cosh{\rho}=z_{0}(\tau),~~t=h_{0}(\tau) \ , \nonumber \\&
  \sin{\theta}=r_{1}(\tau),~~\phi_1=\Phi_{1}(\tau,\sigma)=f_{1}(\tau)+\bar{m}_{1}\sigma \ , \nonumber \\& \cos{\theta}=r_{2}(\tau),~~\phi_2=\Phi_{2}(\tau,\sigma)=f_{2}(\tau)+\bar{m}_{2}\sigma \ .
   \end{eqnarray}Winding numbers $\bar{m}_{a}$'s are assumed along the $\sigma$ direction only to make the time-direction single-valued. With this embedding, the Lagrangian can now be expressed as,
   \begin{equation}
       \begin{split}
           &\mathcal{L}=\frac{\tau_{(m,n)}L^{2}}{2}\left[-\left(\dot{z}_{0}^{2}+z_{0}^{2}\dot{h}_{0}^{2}\right)+\sum_{a=1}^{2}\left(\dot{r}_{a}^{2}+r_{a}^{2}\dot{f}_{a}^{2}-r_{a}^{2}\bar{m}_{a}^{2}\right)\right]\\&+\left(q_{(m,n)}q+\tilde{q}_{(m,n)}\sqrt{1-q^{2}}\right)L^{2}r_{2}^{2}\left(\bar{m}_{2}\dot{f}_{1}-\bar{m}_{1}\dot{f}_{2}\right)-\frac{\Lambda}{2}\left(\sum_{a=1}^{2}r_{a}^{2}-1\right)-\frac{\Tilde{\Lambda}}{2}\left(z_{0}^{2}+1 \right)\ , 
       \end{split}
       \label{pulsatingL}
   \end{equation}where the derivative with respect to $\tau$ is denoted by dots. $\Lambda$ and $\tilde{\Lambda}$ are suitable Lagrange multipliers.
   \subsection{Lagrangian and Hamiltonian formulation}
  Euler-Lagrange equations of motion for $z_{0}$ and $f_{a}$'s can be derived from the Lagrangian (\ref{pulsatingL}) as
   \begin{subequations}
   \begin{align}
       &\ddot{z}_{0}-\frac{C_{0}^{2}}{4z_{0}^{3}}+\tilde{A}z_{0}=0\\&
       \dot{f}_{1}=\frac{C_{1}}{r_{1}^{2}}-\frac{Qr_{2}^{2}\bar{m}_{2}}{r_{1}^{2}}\\&
       \dot{f}_{2}=\frac{C_{2}}{r_{2}^{2}}+Q\bar{m}_{1} \ ,
   \end{align}
   \end{subequations}where ,$\dot{h}_{0}=\frac{C_{0}}{2z_{0}^{2}}$ and $\tilde{A}=\frac{\tilde{\Lambda}}{\tau_{(m,n)}L^{2}}$. Putting the expressions for $\dot{f}_{1}$ and $\dot{f}_{2}$ in terms of $r_{1}$ and $r_{2}$ in the Lagrangian (\ref{pulsatingL}) we achieve,
   \begin{equation}
       \begin{split}
 &\mathcal{L}=\frac{\tau_{(m,n)}L^{2}}{2}\left[-\left(\dot{z}_{0}^{2}+z_{0}^{2}\dot{h}_{0}^{2}\right)+\sum_{a=1}^{2}\left(\dot{r}_{a}^{2}+\frac{\left(C_{a}+Q\bar{m}_{b}r_{2}^{2}\epsilon_{ba}\right)^{2}}{r_{a}^
 {2}}-r_{a}^{2}\bar{m}_{a}^{2}\right)\right]\\&-Q\tau_{(m,n)}L^{2}r_{2}^{2}\left(\frac{\bar{m}_{1}C_{2}}{r_{2}^{2}}-\frac{\bar{m}_{2}C_{1}}{r_{1}^{2}}+\frac{Q\bar{m}_{1}^{2}r_{2}^{2}}{r_{2}^{2}}+\frac{Q\bar{m}_{2}^{2}r_{2}^{2}}{r_{1}^{2}}\right)-\frac{\Lambda}{2}\left(\sum_{a=1}^{2}r_{a}^{2}-1\right)-\frac{\Tilde{\Lambda}}{2}\left(z_{0}^{2}+1 \right) \ .
       \end{split}
   \end{equation}  We can get the equations of motion for $r_{1}$ and $r_{2}$ as
   \begin{equation}
       \ddot{r}_{1}+\frac{\left(C_{1}-Q\bar{m}_{2}r_{2}^{2}\right)^{2}}{r_{1}^
             {3}}+\bar{m}_{1}^{2}r_{1}+\frac{2Q(\bar{m}_{2}C_{1}-Q\bar{m}_{2}r_{2}^{2})r_{2}^{2}}{r_{1}^{3}}+Ar_{1}=0 \ ,
   \end{equation}and
   \begin{equation}
   \begin{split}
         &\ddot{r}_{2}+\frac{Q\bar{m}_{2}r_{2}\left(C_{1}-Q\bar{m}_{2}r_{2}^{2}\right)}{r_{1}^{2}}-\frac{Q\bar{m}_{1}r_{2}\left(C_{2}+Q\bar{m}_{1}r_{2}^{2}\right)}{r_{2}^{2}}+\frac{\left(C_{2}+Q\bar{m}_{1}r_{2}^{2}\right)^{2}}{r_{2}^{3}}+\bar{m}_{2}^{2}r_{2}\\&+2Q\left(Q\bar{m}_{1}^{2}r_{2}-\frac{\bar{m}_{2}C_{1}r_{2}}{r_{1}^{2}}+\frac{2Q\bar{m}_{2}^{2}r_{2}^{3}}{r_{1}^{2}}\right)+Ar_{2}=0 \ .
   \end{split}
\end{equation}These three equations of motion for $z_{0}$, $r_{1}$ and $r_{2}$ can also be obtained from the following Lagrangian 
\begin{equation}
\begin{split}
    L_{NR}=&\frac{1}{2}\left(\dot{z}_{0}^{2}+\dot{r}_{1}^{2}+\dot{r}_{2}^{2}\right)-\frac{1}{2}\frac{\left(C_{1}-Q\bar{m}_{2}r_{2}^{2}\right)^{2}}{r_{1}^{2}}-\frac{1}{2}\frac{\left(C_{2}+Q\bar{m}_{1}r_{2}^{2}\right)^{2}}{r_{2}^{2}}-\frac{C_{0}^{2}}{8z_{0}^{2}}+\frac{1}{2}\left(\bar{m}_{1}^{2}r_{1}^{2}+\bar{m}_{2}^{2}r_{2}^{2}\right)\\&
    +\frac{A}{2}\left(r_{1}^{2}+r_{2}^{2}-1\right)+\frac{\tilde{A}}{2}z_{0}^{2}+\frac{1}{2}Qr_{2}^{2}\left(\frac{\bar{m}_{1}C_{2}}{r_{2}^{2}}-\frac{\bar{m}_{2}C_{1}}{r_{1}^{2}}+\frac{Q\bar{m}_{1}^{2}r_{2}^{2}}{r_{2}^{2}}+\frac{Q\bar{m}_{2}^{2}r_{2}^{2}}{r_{1}^{2}}\right) \ .
    \end{split}
\end{equation}Therefore the Hamiltonian of the system is
\begin{equation}
\begin{split}
        H_{NR}=&\frac{1}{2}\left(\dot{z}_{0}^{2}+\dot{r}_{1}^{2}+\dot{r}_{2}^{2}\right)+\frac{1}{2}\frac{\left(C_{1}-Q\bar{m}_{2}r_{2}^{2}\right)^{2}}{r_{1}^{2}}+\frac{1}{2}\frac{\left(C_{2}+Q\bar{m}_{1}r_{2}^{2}\right)^{2}}{r_{2}^{2}}+\frac{C_{0}^{2}}{8z_{0}^{2}}-\frac{1}{2}\left(\bar{m}_{1}^{2}r_{1}^{2}+\bar{m}_{2}^{2}r_{2}^{2}\right)\\&
    -\frac{A}{2}\left(r_{1}^{2}+r_{2}^{2}-1\right)-\frac{\tilde{A}}{2}z_{0}^{2}-\frac{1}{2}Qr_{2}^{2}\left(\frac{\bar{m}_{1}C_{2}}{r_{2}^{2}}-\frac{\bar{m}_{2}C_{1}}{r_{1}^{2}}+\frac{Q\bar{m}_{1}^{2}r_{2}^{2}}{r_{2}^{2}}+\frac{Q\bar{m}_{2}^{2}r_{2}^{2}}{r_{1}^{2}}\right) \ .
    \end{split}
\end{equation}From these two expressions it is obvious that the forms of both the Lagrangian and Hamiltonian are in agreement with those of the one-dimensional Neumann-Rosochatius system, only with extra terms due to the presence of mixed flux in the background.

The Uhlenbeck integrals of motion may be obtained in a similar way as used in the case of the spinning string ansatz. These yield the expressions as
\begin{equation}
        \begin{split}
          \bar{I_{a}}=&\frac{1}{\bar{m}_{1}^{2}-\bar{m}_{2}^{2}}\left(r_{1}\dot{r}_{2}-\dot{r}_{1}r_{2}\right)^{2}+\frac{2}{\bar{m}_{1}^{2}-\bar{m}_{2}^{2}}\left[\left(\frac{C_{1}-Qr_{2}^{2}\bar{m}_{2}}{r_{1}}\right)^{2}+\left(\frac{C_{2}+Qr_{2}^{2}\bar{m}_{1}}{r_{2}}\right)^{2}\right]\\&-\frac{2}{\bar{m}_{1}^{2}-\bar{m}_{2}^{2}}\left[\left(1+\frac{2Q^{2}r_{2}^{2}}{r_{1}^{2}}\right)\left(\bar{m}_{1}^{2}r_{1}^{2}+\bar{m}_{2}^{2}r_{2}^{2}\right)+2Qr_{2}^{2}\left(\frac{C_{1}\bar{m}_{2}}{r_{1}^{2}}-\frac{C_{2}\bar{m}_{1}}{r_{2}^{2}}\right)\right]\\&+\frac{1}{\bar{m}_{1}^{2}-\bar{m}_{2}^{2}}\left(\frac{C_{1}^{2}}{r_{1}^{2}}+\frac{C_{2}^{2}r_{1}^{2}}{r_{2}^{2}}\right)  \ .
          \label{pulsatingintegral}
        \end{split}
\end{equation} 
The two Virasoro constraints $G_{\tau\tau}+G_{\sigma\sigma}=0$ and $G_{\tau\sigma}=G_{\sigma\tau}=0$ can be calculated as \begin{subequations}
   \begin{align}
       &\dot{r}_{1}^{2}+\dot{r}_{2}^{2}+\dot{f}_{1}^{2}r_{1}^{2}+\dot{f}_{2}^{2}r_{2}^{2}+\bar{m}_{1}^{2}r_{1}^{2}+\bar{m}_{2}^{2}r_{2}^{2}=\dot{z}_{0}^{2}+z_{0}^{2}\dot{h}_{0}^{2}\label{VC1} \ , \\&
       \bar{m}_{1}r_{1}^{2}\dot{f}_{1}+\bar{m}_{2}r_{2}\dot{f}_{2}=0
   \end{align}
\end{subequations}respectively.
\subsection{Solutions for $r_{1}$ and $r_{2}$ and string profile}
Here we use the same procedure for finding the pulsating solutions to this integrable system by choosing ellipsoidal coordinates keeping the conditions as mentioned in section 2.3 intact. In case of pulsating string the ellipsoidal coordinates are taken to be the functions of $\tau$ only and the roots of the equation 
 \begin{equation}
           \frac{r_{1}^{2}}{\zeta-\bar{m}_{1}^{2}}+\frac{r_{2}^{2}}{\zeta-\bar{m}_{2}^{2}}=0 \ ,
           \label{ellipsoidal1}
\end{equation}may be found in terms of $\zeta$. The time derivative of this equation gives 
\begin{equation}
    \left(r_{1}\dot{r}_{2}-\dot{r}_{1}r_{2}\right)^{2}=\frac{\dot{\zeta}^{2}}{4(\bar{m}_{1}^{2}-\zeta)(\zeta-\bar{m}_{2}^{2})} \ .
\end{equation}and also gives $r_{1}^{2}$ and $r_{2}^{2}$ as
\begin{equation}
   r_{1}^{2}=\frac{\bar{m}_{1}^{2}-\zeta}{\bar{m}_{1}^{2}-\bar{m}_{2}^{2}},~~r_{2}^{2}=\frac{\zeta-\bar{m}_{2}^{2}}{\bar{m}_{1}^{2}-\bar{m}_{2}^{2}} \ .
\end{equation} Solving equation (\ref{pulsatingintegral}) for $\dot{\zeta}^{2}$ we can get a similar solution containing 3rd order polynomial exactly like the case for rotating string and it is given as 
\begin{equation}
    \dot{\zeta}^{2}=-4P_{3}(\zeta),~~ P_{3}(\zeta)=2(1-Q)\prod_{i=1}^{3}\left(\zeta-\zeta_{i}\right) \ .
\end{equation} Using the same change of variables as given in equation (\ref{variablechange}) and  with $\eta$ taken to be a function of $\tau$ only we can get 
\begin{equation}
     \eta(\tau)=\hbox{cn}\left(\tau\sqrt{2\left(1-Q\right)\left(\zeta_{3}-\zeta_{1}\right)}+\tau_{0}, k\right) \ 
\end{equation} We may choose integration constant $\tau_{0}$ and the elliptic modulus $k=\frac{\zeta_{3}-\zeta_{2}}{\zeta_{3}-\zeta_{1}}$. From equation (\ref{ellipsoidal1}) we get the expressions for $r_{1}^{2}$ and $r_{2}^{2}$ as 
\begin{equation} r_{1}^{2}(\tau)=\frac{\zeta_{3}-\bar{m}_{1}^{2}}{\bar{m}_{2}^{2}-\bar{m}_{1}^{2}}+\frac{\zeta_{2}-\zeta_{3}}{\bar{m}_{2}^{2}-\bar{m}_{1}^{2}}\hbox{sn}^{2}\left(\tau\sqrt{2\left(1-Q\right)\left(\zeta_{3}-\zeta_{1}\right)}, k\right) \ ,
\label{r1equn}
\end{equation}and
\begin{equation}
    r_{2}^{2}(\tau)=\frac{\bar{m}_{2}^{2}-\zeta_{3}}{\bar{m}_{2}^{2}-\bar{m}_{1}^{2}}-\frac{\zeta_{2}-\zeta_{3}}{\bar{m}_{2}^{2}-\bar{m}_{1}^{2}}\hbox{sn}^{2}\left(\tau\sqrt{2\left(1-Q\right)\left(\zeta_{3}-\zeta_{1}\right)}, k\right) \ .
    \label{r2equn}
\end{equation}Here also the hidden constraint for spherical geometry is satisfied by the expressions (\ref{r1equn}) and (\ref{r2equn}). To restrict the elliptic modulus $k$ within the fundamental domain, i.e,  $0<k<1$, we should choose the order of the $\zeta_{i}$'s as  $\zeta_{1}<\zeta_{2}<\zeta_{3}$. Similar to the case for rotating string, here also $\zeta$ is assumed to cover $\frac{1}{4}$th of the sphere with $r_{i}^{2}\geq0$. This consequently binds the range of $\zeta$ as $\bar{m}_{1}^{2}\leq\zeta\leq\bar{m}_{2}^{2}$ .
\subsection{Conserved Charges and Dispersion Relation}
 The conserved quantities, in the case of pulsating string may be found from the target space Lagrangian in the given background, as follows
\begin{equation}
    E=\frac{\partial \mathcal{L}}{\partial(\partial_{\tau}t)}=-\tau_{(m,n)}L^{2}z_{0}^{2}\dot{h}_{0},~~J_{a}=\frac{\partial \mathcal{L}}{\partial(\partial_{\tau}\phi_{a})}=\tau_{(m,n)}L^{2}\left(r_{a}^{2}\dot{f}_{a}+Qr_{2}^{2}m_{b}\epsilon_{ab}\right) \ .
\end{equation}Here, by taking $z_{0}=1$ we get $\dot{z}_{0}^{2}+z_{0}^{2}\dot{h}_{0}^{2}=\frac{E}{\tau_{(m,n)}L^{2}}$ with $\dot{h}_{0}^{2}=\frac{\mathcal{E}^{2}}{z_{0}^{4}}$. We may express the Virasoro constraint (\ref{VC1}) as a quartic equation of $r_{1}$ by substituting for $\dot{h}_{0}$ and $\dot{f}_{a}$ in terms of conserved energy $\mathcal{E}=\frac{E}{\tau_{(m,n)L^{2}}}$ and angular momenta $\mathcal{J}_{a}=\frac{J_{a}}{\tau_{(m,n)L^{2}}}$. To make it simple, let us consider the case  $\mathcal{J}_{1}=\mathcal{J}_{2}$ and $\bar{m}_{1}=-\bar{m}_{2}$. With the above, we get
\begin{equation}
    r_{1}\dot{r}_{1}=\sqrt{\left[\bar{m}_{1}^{2}(1+Q)-\mathcal{E}^{2}\right]r_{1}^{4}+\left[\mathcal{E}^{2}-\bar{m}_{1}^{2}(1+2Q)+2Q\bar{m}_{1}\mathcal{J}_{1}\right]r_{1}^{2}-(\mathcal{J}_{1}^{2}-\bar{m}_{1}^{2}+2Q\bar{m}_{1}\mathcal{J}_{1})}
\end{equation}It is obvious from the above equation that the expression $\dot{r}_{1}^{2}$ assumes infinite values for both $r_{1}\rightarrow0$ and $r_{1}\rightarrow \infty$ and oscillates in between. Therefore it must have some minimum value which can be derived to be $\left[\mathcal{E}^{2}-\bar{m}_{1}^{2}(1+2Q)+2Q\bar{m}_{1}\mathcal{J}_{1}\right]$ at $r_{1}^{2}=\pm\sqrt{\frac{\mathcal{J}_{1}^{2}-\bar{m}_{1}^{2}+2Q\bar{m}_{1}\mathcal{J}_{1}}{\bar{m}_{1}^{2}(1+Q)-\mathcal{E}^{2}}}$. This yields the quantum oscillation number \cite{Park:2005kt} as 
\begin{equation}
\begin{split}
    &\mathcal{N}=\frac{N}{\tau_{(m,n)L^{2}}}=\oint r_{1}\dot{r}_{1}dr_{1}\\&=\int_{0}^{\sqrt{a_{+}}} dr_{1} \sqrt{\left[\bar{m}_{1}^{2}(1+Q)-\mathcal{E}^{2}\right]r_{1}^{4}+\left[\mathcal{E}^{2}-\bar{m}_{1}^{2}(1+2Q)+2Q\bar{m}_{1}\mathcal{J}_{1}\right]r_{1}^{2}-(\mathcal{J}_{1}^{2}-\bar{m}_{1}^{2}+2Q\bar{m}_{1}\mathcal{J}_{1})}\\& =\int_{0}^{\sqrt{a_{+}}} dr_{1}\sqrt{\left(r_{1}^{2}-a_{-}\right)(a_{+}-r_{1}^{2})} \ ,
    \end{split}
\end{equation}where $a_{\pm}$ are the roots of the biquadratic function under the square root. This quantum number being a large adiabatic quantity determines the string level and it is generally a convenient way to compute the dispersion relation for pulsating string in terms of $\mathcal{N}$ and $\mathcal{J}$ as done in \cite{Minahan:2002rc,Park:2005kt,Pradhan:2013sja}.  Taking partial derivative of the above equation with respect to $m_{1}$ we get
\begin{equation}
    \frac{\partial \mathcal{N}}{\partial\bar{ m}_{1}}=\int_{0}^{\sqrt{a_{+}}} dr_{1}\left[\frac{\bar{m}_{1}(1+Q)r_{1}^{4}-\left[\bar{m}_{1}(1+2Q)-Q\mathcal{J}_{1}\right]r_{1}^{2}+(\bar{m}_{1}-Q\mathcal{J}_{1})}{\sqrt{\left(r_{1}^{2}-a_{-}\right)(a_{+}-r_{1}^{2})}}\right] \ .
    \label{oscillation number}
\end{equation}Expressing the integrals in terms of standard elliptic integrals we are left with
\begin{equation}
    \begin{split}
         \frac{\partial \mathcal{N}}{\partial \bar{m}_{1}}=&\left(\bar{m}_{1}-Q\mathcal{J}_{1}\right)\frac{1}{\sqrt{a_{+}}}\textbf{K}(\epsilon)-\left[\bar{m}_{1}(1+2Q)-Q\mathcal{J}_{1}\right]\sqrt{a_{+}}\left[\textbf{K}(\epsilon)-\textbf{E}(\epsilon)\right]\\&-\frac{\bar{m}_{1}(1+Q)}{3}\sqrt{a_{+}}\left[a_{-}+2\left(a_{+}+a_{-}\right)\textbf{E}(\epsilon)-\left(a_{-}+2a_{+}\right)\textbf{K}(\epsilon)\right] \ ,
    \end{split}
\end{equation}where we have used $\epsilon=\frac{a_{-}}{a_{+}}$. The standard expansions of  elliptic integrals of first kind and second kind are respectively 
\begin{subequations}
  \begin{align}
     &\textbf{K}\left(\epsilon\right)=\frac{\pi}{2}+\frac{\pi\epsilon}{8}+\frac{9\pi\epsilon^{2}}{128}+\frac{25\pi\epsilon^{3}}{512}+\frac{1225\pi\epsilon^{4}}{32768}+ \mathcal{O}[\epsilon]^{5} \ , \\&
     \textbf{E}\left(\epsilon\right)=\frac{\pi}{2}-\frac{\pi\epsilon}{8}-\frac{3\pi\epsilon^{2}}{128}-\frac{5\pi\epsilon^{3}}{512}-\frac{175\pi\epsilon^{4}}{32768}+\mathcal{O}[\epsilon]^{5} \ .
  \end{align} 
\end{subequations}Substituting for these expansions in equation (\ref{oscillation number}) and taking the short string limit, we achieve 
\begin{equation}
    \begin{split}
        &\frac{\partial\mathcal{N}}{\partial \bar{m}_{1}}=\sqrt{2}(\bar{m}_{1}-Q\mathcal{J}_{1})\left[\frac{89\pi}{128}+\frac{123\pi}{128}\frac{Q\mathcal{J}_{1}}{\bar{m}_{1}(1+Q)}+\mathcal{O}\left(\frac{\mathcal{J}_{1}^{2}}{\bar{m}_{1}^{2}}\right)\right]+\\&\frac{1}{\sqrt{2}}\{\bar{m}_{1}(1+2Q)-Q\mathcal{J}_{1}\}\left[\frac{11\pi}{32}+\frac{11\pi}{32}\frac{Q\mathcal{J}_{1}}{\bar{m}_{1}(1+Q)}+\mathcal{O}\left(\frac{\mathcal{J}_{1}^{2}}{\bar{m}_{1}^{2}}\right)\right]\\&+\bar{m}_{1}(1+Q)\left[\frac{19\pi}{48}-\frac{35\pi}{48}\frac{Q\mathcal{J}_{1}}{\bar{m}_{1}(1+Q)}+\mathcal{O}\left(\frac{\mathcal{J}_{1}^{2}}{\bar{m}_{1}^{2}}\right)\right]+\\&\sqrt{2}(\bar{m}_{1}-Q\mathcal{J}_{1})\left[\frac{89\pi}{256}\frac{Q}{\bar{m}_{1}^{2}(1+Q)^{2}}-\frac{267\pi}{128}\frac{Q^{3}\mathcal{J}_{1}}{\bar{m}_{1}^{3}(1+Q)^{4}}+\mathcal{O}\left(\frac{\mathcal{J}_{1}^{2}}{\bar{m}_{1}^{4}}\right)\right]\mathcal{E}^{2}+\\&\frac{1}{\sqrt{2}}\{\bar{m}_{1}(1+2Q)-Q\mathcal{J}_{1}\}\left[\frac{11\pi}{64}\frac{Q}{\bar{m}_{1}^{2}(1+Q)^{2}}+\frac{37\pi}{32}\frac{Q^{2}\mathcal{J}_{1}}{\bar{m}_{1}^{3}(1+Q)^{3}}+\mathcal{O}\left(\frac{\mathcal{J}_{1}^{2}}{\bar{m}_{1}^{4}}\right)\right]\mathcal{E}^{2}\\&+\bar{m}_{1}(1+Q)\left[-\frac{57\pi}{96}\frac{Q}{\bar{m}_{1}^{2}(1+Q)^{2}}-\frac{7\pi}{12}\frac{Q^{2}\mathcal{J}_{1}}{\bar{m}_{1}^{3}(1+Q)^{3}}+\mathcal{O}\left(\frac{\mathcal{J}_{1}^{2}}{\bar{m}_{1}^{4}}\right)\right]\mathcal{E}^{2}\\&+\sqrt{2}(\bar{m}_{1}-Q\mathcal{J}_{1})\left[-\frac{15\pi}{32}\frac{Q^{2}}{\bar{m}_{1}^{4}(1+Q)^{4}}+\frac{45\pi}{128}\frac{Q^{3}\mathcal{J}_{1}}{\bar{m}_{1}^{5}(1+Q)^{5}}+\mathcal{O}\left(\frac{\mathcal{J}_{1}^{2}}{\bar{m}_{1}^{6}}\right)\right]\mathcal{E}^{4}+\\&\frac{1}{\sqrt{2}}\{\bar{m}_{1}(1+2Q)-Q\mathcal{J}_{1}\}\left[\frac{5\pi}{16}\frac{Q^{2}}{\bar{m}_{1}^{4}(1+Q)^{4}}+\frac{9\pi}{32}\frac{Q^{3}\mathcal{J}_{1}}{\bar{m}_{1}^{5}(1+Q)^{5}}+\mathcal{O}\left(\frac{\mathcal{J}_{1}^{2}}{\bar{m}_{1}^{6}}\right)\right]\mathcal{E}^{4}\\&+\bar{m}_{1}(1+Q)\left[\frac{\pi}{12}\frac{Q^{2}}{\bar{m}_{1}^{4}(1+Q)^{4}}-\frac{11\pi}{48}\frac{Q^{3}\mathcal{J}_{1}}{\bar{m}_{1}^{5}(1+Q)^{5}}+\mathcal{O}\left(\frac{\mathcal{J}_{1}^{2}}{\bar{m}_{1}^{6}}\right)\right]\mathcal{E}^{4}+\mathcal{O}\left(\mathcal{E}^{6}\right) \ .
    \end{split}
    \label{pulsating}
   \end{equation}Integrating the equation (\ref{pulsating}) with respect to $\bar{m}_{1}$
and then expanding $\mathcal{E}$ as a function of oscillation number $\mathcal{N}$ and angular momentum $\mathcal{J}_{1}$ for small values of $\mathcal{J}_{1}$ we achieve
\begin{equation}
\begin{split}
    \mathcal{E}^{2}=&\left(25.6704+1.8333\mathcal{N}\right)+\left(24.7749+1.2526\mathcal{N}\right)\mathcal{J}_{1}+\left(6.0259+0.0082\mathcal{N}\right)\mathcal{J}_{1}^{2}\\&-\left(0.311+0.0827\mathcal{N}\right)\mathcal{J}_{1}^{3}+\mathcal{O}[\mathcal{J}_{1}^{4}] \ ,
    \end{split}
    \label{equationE}
\end{equation}where we have taken $Q=1$ and considered the limit for $\bar{m}_{1}\rightarrow 1$. With such limit in hand, small angular momenta, i.e., $\mathcal{J}_{1} \rightarrow 0$ yields from equation (\ref{equationE})
\begin{equation*}
    \mathcal{E}\approx 1.354\sqrt{\mathcal{N}} \ .
\end{equation*}This result resembles the energy-oscillation number relation with small energy limit for fundamental string pulsating in background with pure NSNS flux 
which can be achieved by assuming $m=1, n=0 $ which yields $q=1$ in equation (\ref{fluxexpression}). Also it matches with the energy and oscillation number expansion upto its leading order term in short string limit in \cite{Pradhan:2013sja} for fundamental string pulsating in one plane.
\section{Conclusion}
We have investigated, in this paper, the features of
$AdS_{3}\times S^{3}$ background in the presence of mixed flux in the light of classical integrable model by considering $(m,n)$ string as a natural probe which is a suitable near horizon $SL(2,\mathbb{Z})$ bound state of $NS5$-branes and fundamental strings. We produced a plausible description of the dynamics of $(m,n)$ string rotating and pulsating in our desired background by constructing the well-known 1d Neumann-Rosochatius integrable system. The Lagrangian, Hamiltonian and integrals of motion we computed are similar to those of integrable NR models along with some deformations added as a consequence of the finite flux in the background. It is henceforth a nice understanding of $(m,n)$ string, as a natural probe, credibly supported by integrable deformation of NR mechanical system. We have also elucidated the periodic circular-type string profile for both the cases of closed $(m,n)$ strings rotating and pulsating with two different angular momenta in $\mathbb{R}_{t}\times S^{3}$. The reduction of Uhlenbeck integrals of motion into some first order differential equations has provided such circular-type  solutions of the deformed NR integrable model and this in turn makes solving integrable equations of motion of the system rather simplified. The systematic approach for Neumann-Rosochatius model has digged a concrete way to end up with some general class of dispersion relations among various conserved charges for both the pulsating and rotating $(m,n)$ string. As an immediate solution we have computed the scaling relation for rotating string with constant radii and two equal angular momenta as an expansion of large $J$ for some arbitrary flux parameter. Also the small energy correction to the scaling relation for pulsating string solution has been found out in terms of oscillation number. Since our results show similar form as that of fundamental string, an intriguing idea of a counterpart of our solutions arises in the spirit of Pohlmeyer Reduced theory as achieved with the bosonic part of the action for fundamental string moving in $R\times S^3$ in \cite{Hoare:2013lja}. The WZ interaction term in the $(m,n)$ string action includes an overall flux parameter as $\left(mq+n\sqrt{1-q^2}\right)$ instead of $q$ which leads us to speculate that the Pohlmeyer Reduced theory for $(m,n)$ string in $R_t\times S^3$ might have a mass parameter $\mu$  
deformed accordingly, which for $n=0$ reduces to the case of F1 string. Moreover the fact that  $(m,n)$ string action in $AdS_3\times S^3$ with mixed three form fluxes can be mapped to $(m^{'},n^{'})$ string action in $AdS_3\times S^3$ with pure NSNS two form fluxes stems for the deformation $\mu \rightarrow \mu \left(1-m^{'2}q^2\right)$ in the corresponding Pohlmeyer Reduced model. It would be interesting to study such PR counterparts of string solutions with $(m,n)$ string as a natural probe to generalise the analysis to the Sine-Gordon and Complex Sine-Gordon models. 

Another promising idea arising as a future direction from our study is to construct an NR model for rigidly rotating and pulsating $(m,n)$ probe string in $AdS_3 \times S^3\times S^3\times S^1$ background with various mixed fluxes, which is another classically integrable counterpart of $AdS_3\times S^3\times M^4$ background and has been previously studied by using Yangian hidden symmetry. A general class of string solutions can also be studied by reducing the   $q$-deformed $AdS_{3}\times S^{3}$ background into 1d NR model. 
We wish to return to these problems in future.


\end{document}